\documentstyle[aps,12pt]{revtex}

\begin{document}
\title{Quantum atomistic solid-state theory}
\author{R.J. Radwa\'{n}ski$^{a,b}$ and Z.~Ropka$^b$}
\address{$^a$Center for Solid State Physics, S$^{nt}$ Filip 5, 31-150 Krak%
\'{o}w\\
$^b$Institute of Physics, Pedagogical University, 30-084 Krak\'{o}w, POLAND}
\maketitle

\begin{abstract}
Postulates of the quantum atomistic solid-state theory (QUASST) have been
presented. QUASST\ is an extension of the crystal-field theory and is
applicable to compounds containing open-shell 3d, 4f and 5f atoms. QUASST\
points out the existence in a solid of the atomic-like discrete electronic
structure determined by the crystal-field and spin-orbit interactions. This
approach unifies the description of 3d and 4f ions and allows to bridge the
atomic physics and the solid-state physics.
\end{abstract}

The aim of the quantum atomistic solid-state theory (QUASST) is the
consistent physical model for the understanding of electronic, magnetic
(e-m) and spectroscopic properties of systems containing atoms with an
incomplete electronic shell, namely 4f, 3d and 5f shells. Surely at present
most evidence for the physical adequateness of QUASST exists for rare-earth
systems - we are quite convinced about the substantial applicability of
QUASST to 3d- and 5f-atom compounds. In presenting QUASST we will
concentrate on 4f-atom compounds but the same holds for 5f- and 3d-atom
compounds. Among different properties we can mention the magnetic moment,
the value and its direction at 0 K and at ambient temperatures, temperature
dependence of the heat capacity and the paramagnetic susceptibility, the
anisotropy of magnetic properties. Surely we are interested in the energy
spectrum of available electronic states and their nature. Surely we think
about the formation of the magnetic moment, its spin and orbital
contributions, as well as a symmetry breaking during the formation of the
magnetic state.

QUASST\ accepts a picture called sometimes ionic, but we have to say that,
according to us, this ionic model never was clearly formulated. Even, when
it is sometimes used it is very often misleadingly used, in particular in
case of 3d ions.

The basic ideas of the CEF theory (put forward by Bethe in 1929) - the
existence of the discrete energy states for the paramagnetic open-shell 4f
ions and the importance of the local symmetry - is taken as the main
ingredients in the construction of the more general theory for the solids -
the quantum atomistic solid-state theory. The basic idea of QUASST is that
i) the paramagnetic atoms preserve much of their atomic properties also
being the part of a solid. Otherwords - 4f/3d/5f atoms exist also in a
solid. The novelty of this simple idea can be understood if one remembers
that in the presently-in-fashion modern theories of solids atoms somewhere
disappear. It is clearly visible in the band theories, more often recently
used for the description of systems with open-shell atoms, that yield the
continuous energy spectrum.

We should explain that in different compounds different ionic states of the
paramagnetic atom can be formed. Pr is, for instance, trivalent in Pr$_{2}$O$%
_{3}$, PrAl$_{2}$ and in PrNi$_{5}$ but is tetravalent in PrO$_{2}$. The
different ionic states we consider as different states of the atom. The good
description of PrNi$_{5}$ and ErNi$_{5}$ [1,2], that are intermetallics,
indicates that there exists the Pr$^{3+}$ and Er$^{3+}$ ions despite of the
conduction band electrons. It means, that we can have the clearly defined
ionic state, with the well-defined integer number of 4f electrons, in an
intermetallic compound. Then, apart of the description of the Pr$^{3+}$ ions
we should discuss properties of conduction electrons. In RENi$_{5}$ their
contribution to the heat capacity and to the paramagnetic susceptibility is
small and is the trivial function of temperature. Similarly the Co atom in
CoO is in the divalent state but it is in the trivalent state in LaCoO$_{3}$.

By the preservation of the atomic structure we understand that the Pr$^{3+}$
ion, for instance, has the atomic-like 4f$^{2}$ system (other electrons are
in close shells). It means that the intra-atomic interactions are strong
enough to preserve this highly-correlated electronic system also when atom
is the part of a solid and undergoes interactions with surroundings.
Similarly six 3d electrons of the Co$^{3+}$ ion form the highly-correlated 3d%
$^{6}$ electronic system. Thus ii) n 4f/3d/5f electrons form the
highly-correlated 4f$^{n}$/3d$^{n}$/5f$^{n}$ electronic system. Subsequently
we have the term and multiplet structure known from the atomic physics. The
intra-atomic correlations lead to iii) the ground term of the 4f$^{n}$/3d$%
^{n}$/5f$^{n}$ system with the resultant S and L given by Hund's rules. Then
we work in the (2S+1)(L+1) space. In case of 4f and 5f ions this space is
further reduced (by the existence of the large spin-orbit coupling) to the
(2J+1) space, where J of the lowest multiplet is given by the 3$^{rd}$
Hund's rule.

QUASST points out iv) the importance of the intra-atomic spin-orbit (s-o)
coupling. So far the 4f ions are described within the large s-o limit
whereas the s-o coupling has been largely ignored for 3d ions. There is the
time to take into account the finite, though quite large, for the s-o
coupling in description of rare-earth ions and the non-zero, though
relatively weak, s-o coupling for the 3d ions. It turns out that the large
s-o limit used in the description of rare-earth ions is quite physically
adequate - it means that taking into account the finite s-o coupling does
not introduce the revolution in the low-energy part of the electronic
structure, but allows to explain, for instance, the appearance of the higher
multiplets at the finite energies (the excited multiplet for the Sm$^{3+}$
ion lies at 120 meV only as detected by inelastic-neutron-scattering
experiments of Furrer et al. [3]). However, in case of the 3d ions taking
into account the s-o coupling, even of the small value, causes the
completely new low-energy electronic structure [4] and dramatically changes
the shape of eigenfunctions in comparison to the situation with $\lambda
_{s-o}$ = 0. In fact, the smaller value of the s-o coupling the lower energy
scale in the electronic structure appears and the lower temperature range,
where anomalies of electronic and magnetic properties appear.

QUASST points out v) the existence of the discrete electronic structure
associated with the atomic-like states of the 4f$^n$/3d$^n$/5f$^n$ systems
in a solid.

QUASST points out vi) the importance of the atomic scale symmetry on the
electronic structure, in particular on the realized ground state and its
magnetic moment, both the value and the direction. This electronic structure
determines the electronic and magnetic properties macroscopically observed
both at zero temperature as well as at ambient temperatures.

QUASST points out vii) the strong interplay of the magnetic state of a
paramagnetic ion and the symmetry of the electric field produced by
surrounding charges. The electric field produced by surrounding charges
within the CEF community is known as the crystal field. The well-known
example is the Pr$^{3+}$ ion in PrNi$_5$. It is the local symmetry of the
crystal field that produces the non-magnetic state of the paramagnetic ion.
Within the CEF community this effect is discussed in the connection to the
Kramers and the Jahn-Teller theorem. The CEF interactions can substantially
reduce the local magnetic moment also in case of the Kramers ions. It has
been found that the hexagonal symmetry CEF can produce a non-magnetic
Kramers-doublet ground state for the highly-correlated f$^3$ system [5].
Recently we have shown [6] that the long-time intriguing non-magnetic state
of LaCoO$_3$ is due to the non-magnetic state of the Co$^{3+}$ ion. This
non-magnetic state is produced in the atomic scale by the trigonal off-cubic
distortion provided the intra-atomic spin-orbit coupling is taken into
account. It has been also shown that viii) for the 3d-ion compounds the
Jahn-Teller theorem has to be considered within the spin-orbital space in
contrary to the orbital-only space considered so far. Such the treatment
makes unification with the rare-earth ions where we customarily consider the
Jahn-Teller theorem within the spin-orbital space. In fact, such the
treatment came to the rare-earth magnetism in a very natural way with the
acceptance after the works of Hund (1925) and Van Vleck (1929, 1932) that J
is the good quantum number for 4f ions. Simultaneously the works of Van
Vleck (1932), Schlapp and Penney (1932) have introduced the quenching of the
orbital moment in the 3d ions that gave a start for the (erroneous)
consideration of 3d-ion magnetism as related with the spin-only moment.

QUASST, providing the discrete electronic structure, ix) enables the
calculation of whole thermodynamics. The obtained temperature dependences of
the heat capacity and the paramagnetic susceptibility are in remarkably good
agreement with experimental data for continuously increasing great number of
compounds [1, 2, 7-12].

QUASST points out x) the multipolar character of the electric field existing
in a solid. The higher-order CEF interactions are very important.
Higher-order CEF\ parameters reflect multipolar charge interactions. The
parameters B$_2^n$, B$_4^n$ and B$_6^n$, for instance, are associated with
the quadrupolar, octupolar and hexadekapolar interactions, respectively, and
all of them have enormous influence on the realized electronic structure.
Note, that their influence cannot be treated as the subsequent expansion
terms.

QUASST\ xi) distinguishes the atomic-like properties of the single ion and
the macroscopic properties. In the simplest case the molar heat capacity is
obtained from the single-ion heat capacity by multiplying by the Avogadro
number. The same holds for the paramagnetic susceptibility and the ordered
magnetic moment. This is justified only in case of the simplest structures.
In general, there can be in the crystal structure a few non-equivalent sites
with the different lattice point symmetry. However, the further site
differentiation can occur by the symmetry of the local crystal field. The
principal axis of the electric field gradient can differ from site to site
forming, for instance, the zig-zag structure. As the symmetry can be very
low the principal axis of the quadrupolar (B$_2^m$), octupolar (B$_4^m$) and
hexadekapolar (B$_6^m$) interactions is necessary to consider too. It makes
the direct correlation between the atomic-like properties and macroscopic
properties not always straightford.

QUASST postulates xii) the lowering symmetry with the lowering temperature
as much as possible as the general rule (the extension of the Jahn-Teller
theorem). It causes that with the lowering temperatures we should take into
account more and more inequivalent sites what makes the CEF and QUASST
calculations much more troublesome.

xiii) The magnetic state is related with the time-reversal symmetry (TRS)
breaking that can be traced by the TRS breaking at the atomic scale as the
Zeeman-like effect [1, 2, 8-10]. The nicest illustration for it is the
splitting of the Kramers doublet ground state. The formation of the magnetic
state is somehow forced by the lowering-energy demand - this lowering energy
is nicely visible even in the atomic scale in case the Kramers doublet
ground state. In case of a non-Kramers system the existence of a
closely-lying localized state helps in the formation of the magnetic state.

xiv) Energies involved in the formation of the magnetic state are relatively
weak. For NiO with T$_N$ of 525 K the energy gain amounts to 3.25 kJ/mol
(=33.5 meV/ion) [11]. It is much smaller than the Stoner splitting I, that
is of order of 0.6-1.2 eV.

xv) The molecular field B$_{mol}$ is quite small. In case of NiO B$_{mol}$
amounts to 510 T [11] and the effect of this field is visible in the
atomic-like discrete electronic structure.

QUASST\ makes the unification in the theoretical description of the 3d and
4f ions.

QUASST\ makes the unification in the theoretical description of the 5f- and
4f- ion compounds. Very good description of electronic, magnetic and
spectroscopic properties of UPd$_{2}$Al$_{3}$ [8], UGa$_{2}$ [9], NpGa$_{2}$
[10] and NpPd$_{2}$Al$_{3}$ have been described within the CEF theory with
the trivalent state of the actinide ions. The remarkably good applicability
is really surprise knowing that all of these compounds are metallic.
Moreover, UPd$_{2}$Al$_{3}$ exhibits at low temperatures superconductivity
and the heavy-fermion behavior.

QUASST starts the discussion of the correlation among the d and f electrons
from the highly-correlated limit.

QUASST starts the description of a solid from the description of the
involved atoms.

QUASST\ bridges the atomic physics and the solid-state physics.

One can ask: ''Is this atomic idea a new one in the solid-state physics?''
Yes and no. No, as most of experimentalists naturally discuss their results
in terms of local properties. Yes, as according to our knowledge noone has
been able to resist to presently-in-fashion solid-state physics theories
that simply ignore the existence of the atom in the solid arguing that the
solid is so many-body object and that there are so strong intersite
correlations that the individuality of atoms is lost. In the standard
band-structure calculations the f and d electrons are taken as itinerant
forming a band. In the band there is a continuum of the energy states within
1-5 eV. In our model there are discrete states with energy separations even
less than 1000 times smaller (1 meV, but 0 in case of Kramers ions). No, as
there are some text books written about the crystal field, let mention a
book of Abragam and Bleaney [13] or Ballhausen [14]. Yes, as they applied
the CEF\ approach to some diluted 3d systems, not to the concentrated ones.
Yes, as they have not been consequent enough and by discussing different
crystal-field approaches (weak, strong, ...) with further concepts (e.g.
low- and high-spin states) they largely washed up the original idea. Please
note that in the strong CEF\ approach the n 3d electrons are treated as
largely independent, i.e. they do not form the highly-correlated 3d$^n$
system in contrary to the present model. Our approach corresponds to the
weak crystal-field approach, but we point out the fundamental importance of
the intra-atomic spin-orbit coupling, despite of its relative weakness for
the 3d ions. Also yes, as at present this atomic-like picture is enormously
prohibited in the leading physical journals and a little is said about the
discrete states in the magnetic and strongly-correlated electron system
conferences.

We would like to add, preceding unfounded critics, that we do not claim that
everything can be explained only by single atoms but our point is that the
proper, i.e. physically adequate starting point for the discussion of
properties of the solid containing the open-shell atoms is the consideration
of its atomic states. Our numerous computer experiments point out that e.g.
the orbital moment has to be unquenched in the solid-state physics of 3d-ion
containing compounds and our approach enables it. For instance, we have
derived the orbital moment in NiO to be 0.54 $\mu _B$ what amounts to 20 \%
of the total moment (2.53 $\mu _B$) [11]. Moreover, one should not consider
our approach as the treatment of an isolated atom - we start the discussion
of NiO from the consideration of the cation octahedra NiO$_6$ (more exactly
- the Ni$^{2+}$ ion in the octahedral crystal field). The whole NaCl
structure of NiO is built up from the edge sharing cation octahedra. The
perovskite structure, for instance, is built up from the corner (and the
edge) sharing cation octahedra along the c direction (in the a-b plane).
Thus, such octahedra cover the whole macroscopic sample provided the perfect
translational symmetry. The CEF\ parameters contain information about the
interaction of the single ion with the whole charge surroundings. Our
approach is in agreement with the general conviction about the importance of
the electron correlations in description of open-shell compounds - in our
approach we start from the very highly-correlated limit in contrary to a
weak correlation limit of the LSDA\ approach.

In conclusion, on basis of the extended analysis of experimental results for
the great number of compounds containing 4f, 5f and 3d open-shell atoms we
have developed the quantum atomistic solid-state theory. This theory points
out the existence of the discrete electronic structure associated with
atomic-like states of the involved 4f, 3d, 5f atoms. The existence of such
the structure causes dramatic changes of the low-temperature electronic and
magnetic properties like the formation (or not) of the local magnetic moment
and its long-range magnetic order, temperature dependence of the magnetic
susceptibility and of the heat capacity and according to us most of the
experimentally observed anomalies originate from these discrete energy
states. For the better illustration of our point of view the reader is asked
to look into recent Phys.Rev.Lett. papers. In Ref. 15 authors, considering
states of two 3d electrons of the V$^{3+}$ ion in V$_{2}$O$_{3}$, came out
with the continuum electronic structure spread over 2.5 eV (Figs 2 and 3).
In Ref. 16 the continuum electronic structure for six 3d electrons in FeO
spreads over 8 eV (Fig. 8). In Refs 15-17 the orbital moment and the
spin-orbit coupling is completely ignored. By this Letter we put the
conjecture that in these cases the d electrons form the crystal-field
discrete energy states with the importance of the s-o coupling. In FeO, in
the paramagnetic state, a quite similar structure to that presented in Refs
4 and 6 is realized. We are convinced that the publishing of our paper
enables the open scientific discussion on the magnetism and the electronic
structure of 3d/4f/5f-atom containing compounds and we are ready for this
discussion.

\end{document}